\documentclass[twocolumn]{aastex63}
\usepackage{amsmath}
\usepackage{mathtools}
\usepackage{amssymb}
\usepackage{lineno}
\def\be{\begin{eqnarray}}
\def\ee{\end{eqnarray}}

\shorttitle{Discriminant Properties in FRBs}
\shortauthors{Zhong et al.}

\begin{document}

\title{Can a Single Population Account for the Discriminant Properties in Fast Radio Bursts?}
\author[0000-0002-1766-6947]{Shu-Qing Zhong}
\affil{Department of Astronomy, School of Physical Sciences, University of Science and Technology of China, Hefei 230026, People’s Republic of China; sqzhong@ustc.edu.cn, daizg@ustc.edu.cn}
\author[0000-0003-0055-5287]{Wen-Jin Xie}
\affil{Université de Paris, CEA Paris-Saclay, IRFU/DAp-AIM, F-91191 Gif-sur-Yvette, France}
\author[0000-0003-0471-365X]{Can-Min Deng}
\affil{Guangxi Key Laboratory for Relativistic Astrophysics, Department of Physics, Guangxi University, Nanning 530004, People’s Republic of China; dengcm@gxu.edu.cn}
\author[0000-0002-8391-5980]{Long Li}
\affil{School of Astronomy and Space Science, Nanjing University, Nanjing 210023, People’s Republic of China}
\author[0000-0002-7835-8585]{Zi-Gao Dai}
\affil{Department of Astronomy, School of Physical Sciences, University of Science and Technology of China, Hefei 230026, People’s Republic of China; sqzhong@ustc.edu.cn, daizg@ustc.edu.cn}
\affil{School of Astronomy and Space Science, Nanjing University, Nanjing 210023, People’s Republic of China}
\author[0000-0001-6863-5369]{Hai-Ming Zhang}
\affil{School of Astronomy and Space Science, Nanjing University, Nanjing 210023, People’s Republic of China}

\begin{abstract} 
It is still a highly debated question as to whether fast radio bursts (FRBs) are classified into one or two populations.
To probe this question, we perform a statistical analysis using the first Canadian Hydrogen Intensity Mapping 
Experiment Fast Radio Burst (CHIME/FRB) catalog 
and identify a few discriminant properties between repeating and non-repeating FRBs such as the repetition rate,
duration, bandwidth, spectral index, peak luminosity, and potential peak frequency.
If repeating and non-repeating FRBs belong to one population, their distribution distinctions for the repetition rate and duration
can be explained by the selection effect due to the beamed emission as in \cite{con20}.
However, we obtain that the distribution distinctions for the spectral index and potentially the peak frequency cannot 
be explained by the beamed emission within the framework of
either the coherent curvature radiation or synchrotron maser emission.
This indicates that there could be two populations.
We further discuss three possible scenarios for the required two populations.
\end{abstract}


\section{Introduction}
\label{sec:introduction}
Fast radio bursts (FRBs) are luminous, millisecond, and cosmological radio flashes \citep{katz18,pop18,cor19,pet19,zhang20,lyub21,xiao21}.
A consensus had not been reached that at least some FRBs originate from magnetars
until FRB 20200428 and its associated X-ray burst, both from the Galactic magnetar SGR 1935+2154, were discovered
 \citep{chime20,boch20,mere20,lick21,tav21,rid21}.

Rapid progress in this field has been made though, there are still many questions such as the origin, mechanism, and classification of FRBs. One highly debated question is whether repeating and non-repeating FRBs are two different populations \citep{pal18,caleb19,ai21}. Observationally, there are indeed a few discriminant properties such as the duration and bandwidth between repeating and non-repeating FRBs reported by the \cite{chime19}, \cite{fon20}, and \cite{ple21} by using the first CHIME/FRB catalog \citep{chime21}, as well as the obvious but easily neglected observed repetition rate between them. However, whether these observational discriminant properties stem from two different populations 
or just one population but due to a selection effect is still unknown. 

\cite{con20} suggested that the striking discrepancy in duration and observed repetition distributions between repeating and non-repeating FRBs 
can result from the selection effect due to a beamed emission\footnote{The invoking of a varying power-law index (may lie with intrinsic differences
in the central engines or environments) of the frequency drift rate in the time-frequency structure of FRBs could generate the dichotomy between burst duration
and bandwidth most recently \citep{met22}.}.
For the distribution discrepancies in other properties between repeating and non-repeating FRBs, we explore whether they can also arise from the beamed emission within the framework of either the coherent curvature radiation or synchrotron maser emission, and further discuss whether it requires two different populations to give rise to these discrepancies. The structure of this paper is organized as follows. A few discriminant properties between repeating and non-repeating FRBs are outlined in Section \ref{sec:properties}. Whether these distinctions can be explained by the beamed emission is presented in Section \ref{sec:explanation}. Section \ref{sec:discussion} presents a discussion of three possible scenarios for the required two populations. Our conclusions are presented in Section \ref{sec:conclusions}.

\section{Discriminant Properties between Repeating and Non-repeating FRBs}
\label{sec:properties}
To make the one-off events and repeaters used for statistical analysis have selection biases that are as identical
as possible, one should
(1) use the first-detected repeater events for each repeating source, 
and (2) cut the events with signal-to-noise ratio $<$ 12, 
those events having an observed dispersion measure (DM) $<$ 1.5 max(DM$_{\rm NE2001}$, DM$_{\rm YMW16}$)\footnote{DM$_{\rm NE2001}$ (DM$_{\rm YMW16}$) is the Milky Way ISM contribution to the observed DM calculated by the model NE2001 (YMW16) as in \cite{cor02} and \citep{yao17}.}, 
and those detected in the far side-lobes \citep{chime21}.
From the statistical analysis for those events that fulfill the above requirements in the first CHIME/FRB catalog, 
a discriminant property like the peak luminosity can be found, and other discriminant properties such as the duration, bandwidth,
spectral index, and potential peak frequency reported by \cite{ple21} can be picked up. 
We provide an overview of them as follows.

\subsection{Duration and Bandwidth}
\label{subsec:duration}
Previously, the \cite{chime19} and \cite{fon20} found that non-repeating FRBs are typically shorter in duration ($\tau$) than those bursts from repeating sources. \cite{fon20} also found that nearly all bursts from repeating sources possess small emission bandwidths ($\Delta\nu_{\rm b}$) $100-200$ MHz, whereas a large fraction of non-repeating FRBs span the full CHIME band of 400$-$800 MHz. More recently, 
\cite{ple21} confirmed these statistical results using the first CHIME/FRB catalog \citep{chime21}.

\begin{figure*}
	\includegraphics[scale=0.35]{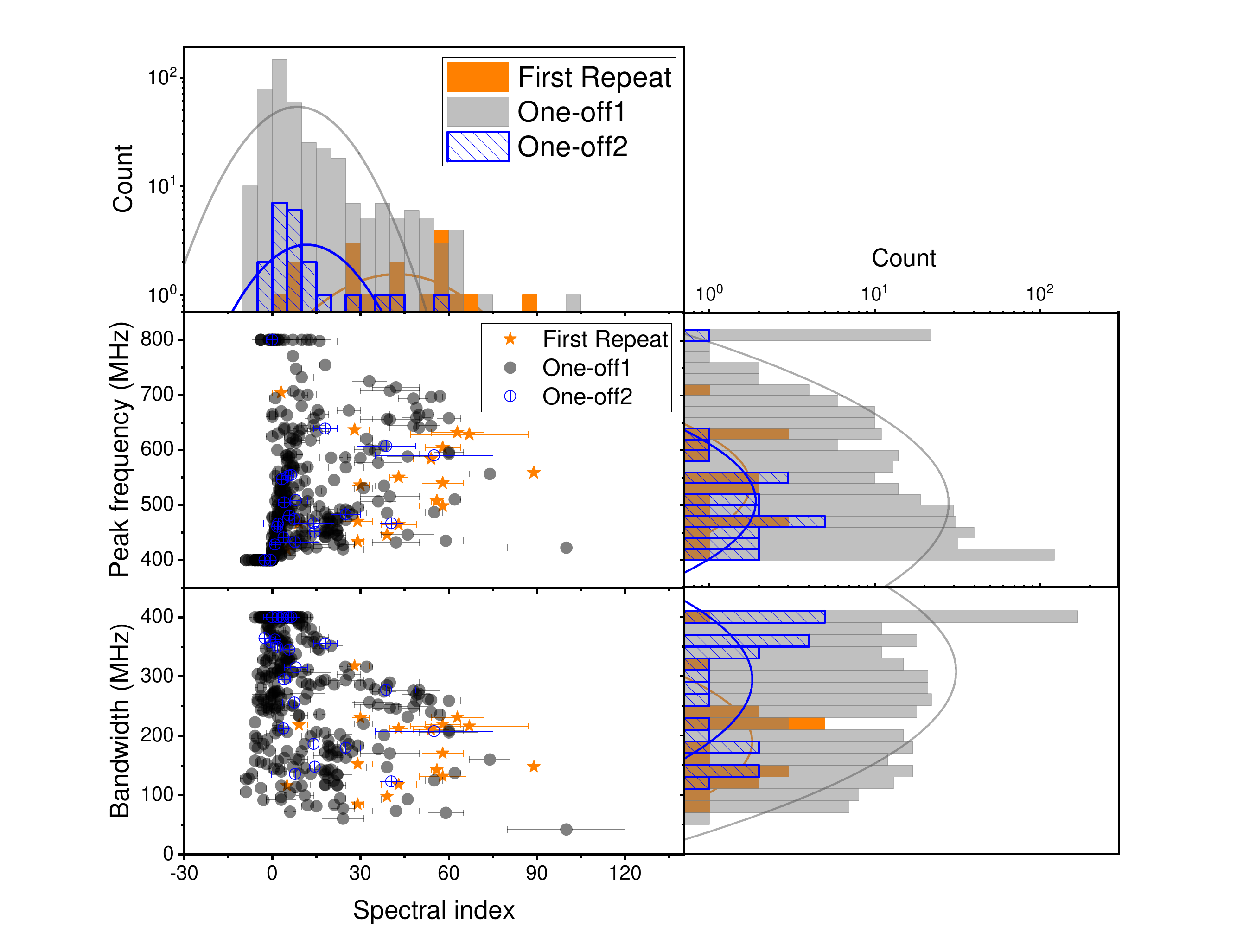}
	\includegraphics[scale=0.35]{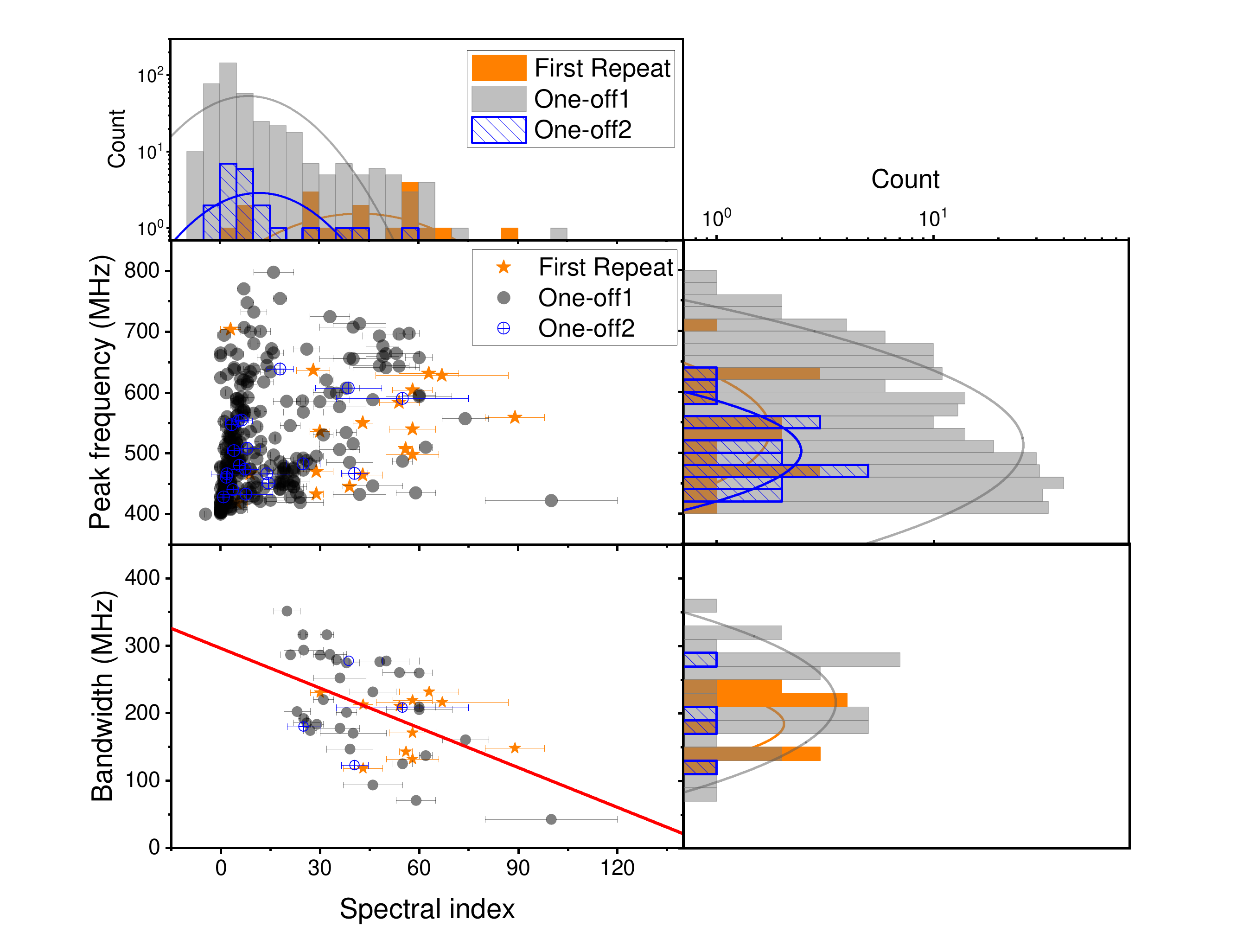}
	\includegraphics[scale=0.35]{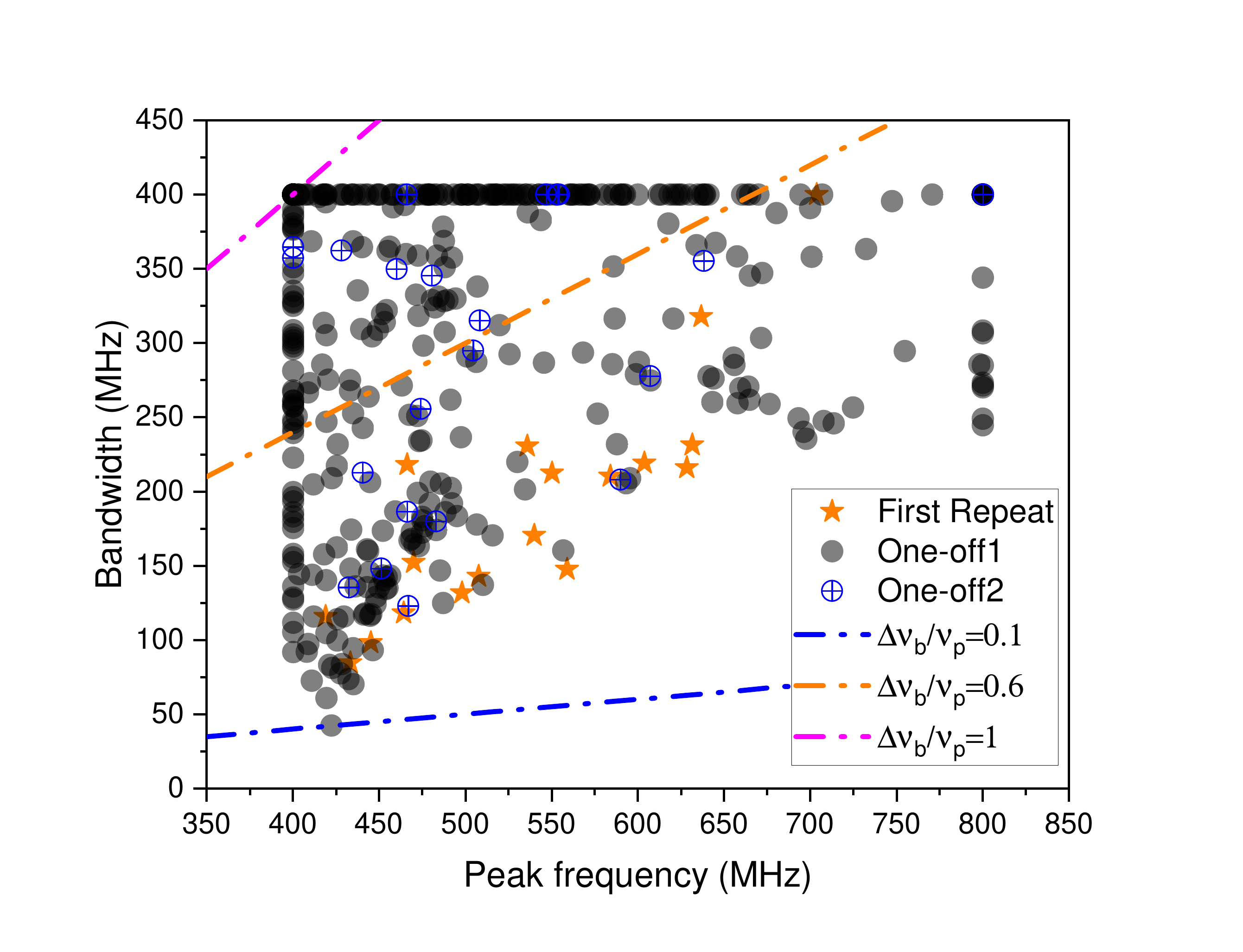}
	\includegraphics[scale=0.35]{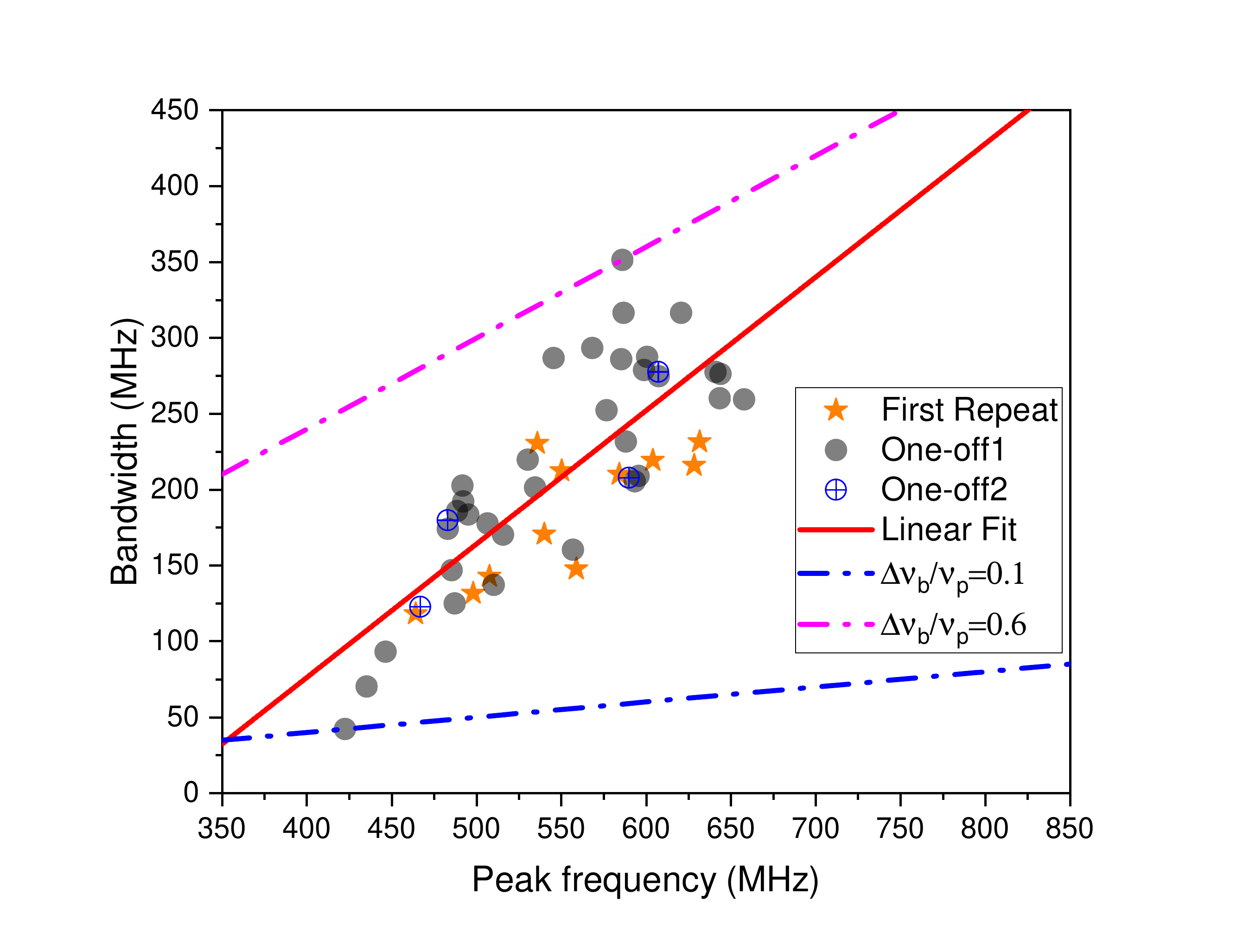}
	\caption{Upper panels: scatter maps and distributions of the spectral index, peak frequency, and bandwidth for the complete samples (left) and incomplete samples (right), including three parts: the first-detected events of repeaters (first repeat sample), 
			single-pulse one-off events (one-off1 sample), and multi-pulse one-off events (one-off2 sample).
		Lower panels: scatter maps between the peak frequency and bandwidth for the complete samples (left) and the incomplete samples (right). The complete and incomplete samples are referred in Footnote \ref{fn:samples}.}
	\label{fig:sp_idx}
\end{figure*}

\begin{deluxetable*}{cccccccccccccc}
	\centering
	\tablewidth{0pt}
	\tabletypesize{\small}
	\tablecaption{Comparison of the Distribution between the First-detected Events of Repeaters and One-off Events}
	\tablehead{
		\colhead{Complete Samples$^a$} &
		\colhead{Bandwidth (MHz)$^b$} &
		\colhead{Spectral Index$^b$} &
		\colhead{Peak Frequency (MHz)$^b$} &
		\colhead{log($L_{\rm p}$/erg s$^{-1}$)$^b$} &
		\colhead{log($E$/erg)$^b$} 
	}
	\startdata
	\object{First repeat} &177$\pm$62&45$\pm$22&528$\pm$73&41.7$\pm$1.3&39.5$\pm$1.4 \\
	\object{One-off1} &308$\pm$103&8.5$\pm$15.0&503$\pm$114&42.6$\pm$1.1&40.0$\pm$1.1 \\
	\object{One-off2} &294$\pm$97&12$\pm$15&507$\pm$92&42.2$\pm$0.9&39.8$\pm$0.9 \\
	\object{$p_{\rm KS}^c$}&1e-7, 1.6e-3, 0.15&4e-10, 4e-5, 0.12&0.041, 0.44, 0.13&3.7e-3, 0.61, 0.06&0.12, 0.47, 0.22 \\
	\object{$p_{\rm AD}^c$}&$<$1e-3, 2.4e-3, 0.17&$<$1e-3, $<$1e-3, 0.11&9.4e-3, $>$0.25, 0.15&4.8e-3, $>$0.25, 0.13&0.15, $>$0.25, $>$0.25 \\
	\hline
	\object{Incomplete Samples$^a$} &&&&& \\
	\hline
	\object{First repeat} &185$\pm$43&45$\pm$22&528$\pm$73&41.7$\pm$1.3&39.5$\pm$1.4 \\
	\object{One-off1} &217$\pm$75&8.5$\pm$15.0&512$\pm$89&42.6$\pm$1.1&40.0$\pm$1.0 \\
	\object{One-off2} &197$\pm$65&12$\pm$15&503$\pm$62&42.2$\pm$0.9&39.8$\pm$0.9 \\
	\object{$p_{\rm KS}^c$}&0.055, 0.89, 0.88&4e-10, 4e-5, 0.12&0.25, 0.53, 0.70&1.6e-3, 0.72, 0.011&0.091, 0.72, 0.34 \\
	\object{$p_{\rm AD}^c$}&0.12, $>$0.25, $>$0.25, &$<$1e-3, $<$1e-3, 0.11&0.22, $>$0.25, $>$0.25&3.2e-3, $>$0.25, 0.065&0.111, $>$0.25, $>$0.25 \\
	\enddata
	\tablenotetext{a}{The complete samples are the ones in which the bursts contain all quantity values, while the incomplete samples are the ones in which the values of the highest frequency, lowest frequency, or peak frequency capped at the top or bottom of the CHIME band of the bursts have been removed, and the relevant values of quantities such as the bandwidth, peak luminosity, and energy calculated from the values of the highest frequency, lowest frequency, or peak frequency have also been removed.}
		\tablenotetext{b}{The results for these quantities are all obtained by fitting their distributions with a Gaussian function.}
		\tablenotetext{c}{The three values are respectively the $p$-values for the first repeat and single-pulse one-off event (one-off1) samples, the first repeat and multi-pulse one-off event (one-off2) samples, and the one-off1 and one-off2 samples, by using the K-S and AD tests encoded in the Python package SciPy \citep{vir20}.}
	\label{tab:results}
\end{deluxetable*}

\subsection{Spectral Index and Peak Frequency}
\label{subsec:sp_idx}
The spectral index ($\gamma$), peak frequency ($\nu_{\rm p}$), and bandwidth are all obtained from 
the fitting and evaluation of the spectral properties of the bursts by using the empirical functional form (1)
with a spectral running term $r$ as in \cite{ple21}. Through the spectral running, \cite{ple21} found that example spectra 
for possible combinations of $\gamma$ and $r$
can be obtained, and the bandwidth and peak frequency are also found directly 
by evaluating the functional form. As a result, the spectral index and running term are expected to be coupled from the spectral running, 
including the bandwidth and peak frequency, 
as they are just two different ways of parameterizing the spectral properties of the bursts with two parameters.
These couplings exist, as shown by the negative correlation tendency between the spectral index and bandwidth 
and the positive correlation tendency between the peak frequency and bandwidth, respectively, in the plotting 
in the upper right and lower right panels of Figure \ref{fig:sp_idx} 
for the incomplete samples\footnote{Note that the bandwidth here is calculated by the highest frequency 
minus the lowest frequency from the first CHIME/FRB catalog. For a large portion of bursts, especially the one-off events, 
they have the highest frequency capped at the top of the CHIME band and/or the lowest frequency capped at the bottom of the CHIME band. 
For these bursts, their highest frequencies should be generally higher than the top of the CHIME band, 
and/or the lowest frequencies lower than the bottom. 
Thus, their real bandwidths should be broader than the calculated values. Similarly, 
for a part of bursts whose peak frequencies capped at the top or the bottom of the CHIME band, 
their real peak frequencies should not be the limits of the CHIME band. Therefore, 
we treat the incomplete samples as being the ones in which the values of the highest frequency, 
lowest frequency, or peak frequency capped at the top or bottom of the CHIME band bursts have been removed, 
and also the relevant values of quantities such as the bandwidth, peak luminosity, 
and energy calculated from	the values of the highest frequency, lowest frequency, or peak frequency have been removed. 
While the complete samples are the ones in which the bursts contain all quantity values, including those obtained from 
the highest frequency, lowest frequency, or peak frequency capped at the CHIME band.\label{fn:samples}} 
only, containing the well-constrained quantity values, 
though they are not presented in the upper left and lower left panels for the complete samples. 
The distribution distinctions for those properties between the repeaters and one-off events also support such couplings.
Taking the spectral index as an example, the one-off events 
(either the single-pulse sample: one-off1, or the multi-pulse sample: one-off2) 
tend to have a relatively small index
(either negative or positive: falling or rising), corresponding to a relatively flat spectrum,
while the repeaters have a very large positive index and no one negative index, relevant to 
an extremely steep rising spectrum, as shown by the Gaussian function fitting results listed in Table \ref{tab:results}. 
Correspondingly, the one-off events on average have a broader bandwidth and a lower peak frequency than the repeaters, 
either for the complete or incomplete samples.
Moreover, the Kolmogorov-Smirnov (KS) and Anderson-Darling (AD) tests of the distributions of the bandwidth, spectral index, 
and peak frequency are also given in Table \ref{tab:results}.
For the spectral index, its $p$-values among the first repeat, one-off1, and one-off2 samples
are corresponding to those for the bandwidth from the complete samples. 
For the peak frequency, its $p$-values ($p_{\rm AD}=0.0094$, $p_{\rm KS}=0.041$) 
show that the evidence of a distribution difference between 
the one-off1 and first repeat samples in the complete samples is marginal but noteworthy.
These $p$-values also seem to correspond to those of the bandwidth between 
the one-off1 and first repeat samples.
It is worth noting that the peak frequencies for the first repeat and one-off2 samples 
cannot be considered to originate from different underlying distributions from the $p$-values.
This result is different from that of the bandwidth and spectral index for the first repeat and one-off2 samples in the complete samples.
Note that for each multi-pulse burst either in the first repeat or one-off2 samples, 
its bandwidth, spectral index, and peak frequency are obtained from an average of its sub-pulses.

\subsection{Peak Luminosity and Energy}
\label{subsec:luminosity}
\begin{figure}
	\includegraphics[scale=0.35]{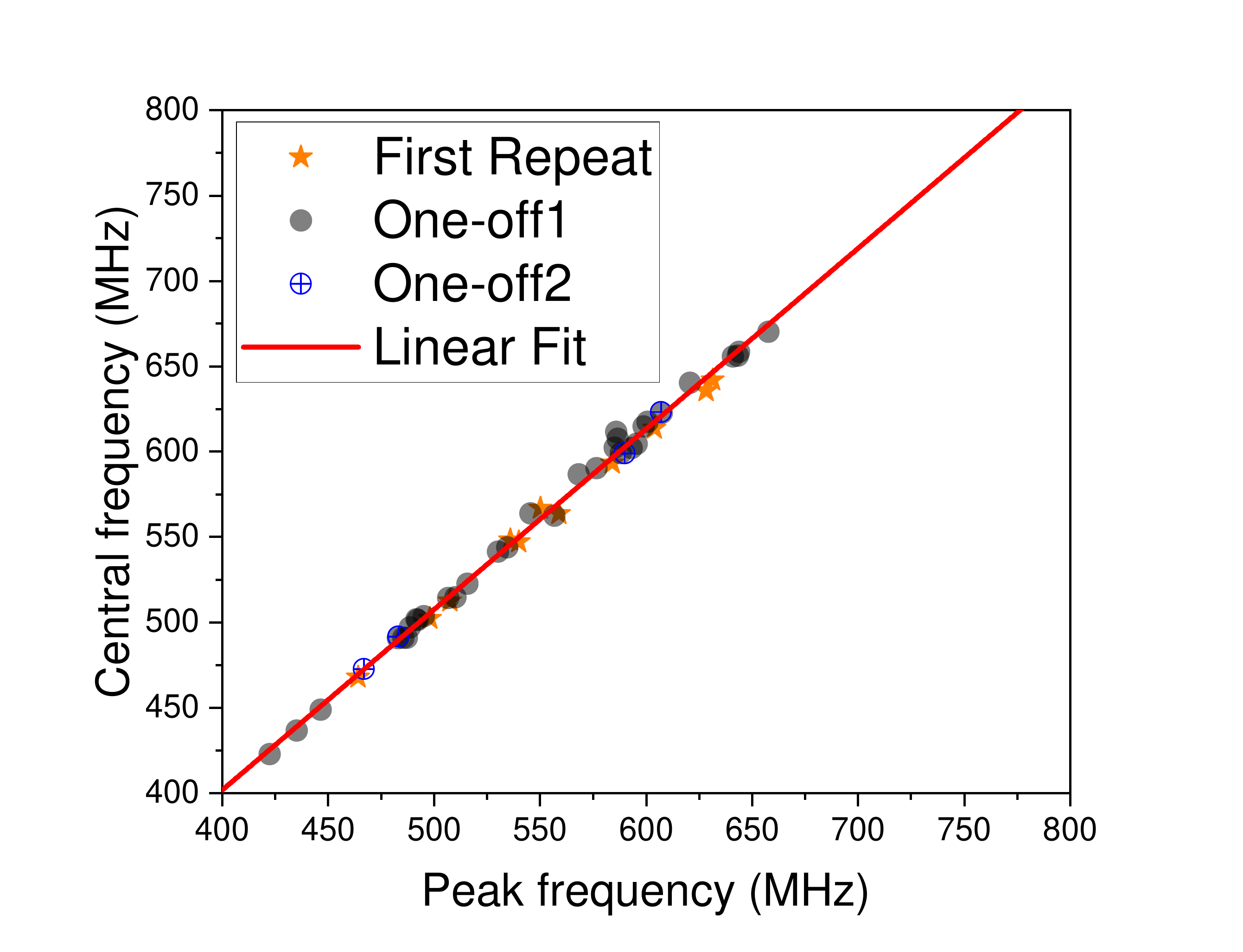}
	\caption{Scatter map comparing the well-estimated central frequencies and well-determined peak frequencies for the first repeat, one-off1, and one-off2 samples.}
	\label{fig:pcfreq}
\end{figure}

\begin{figure}
	\includegraphics[scale=0.35]{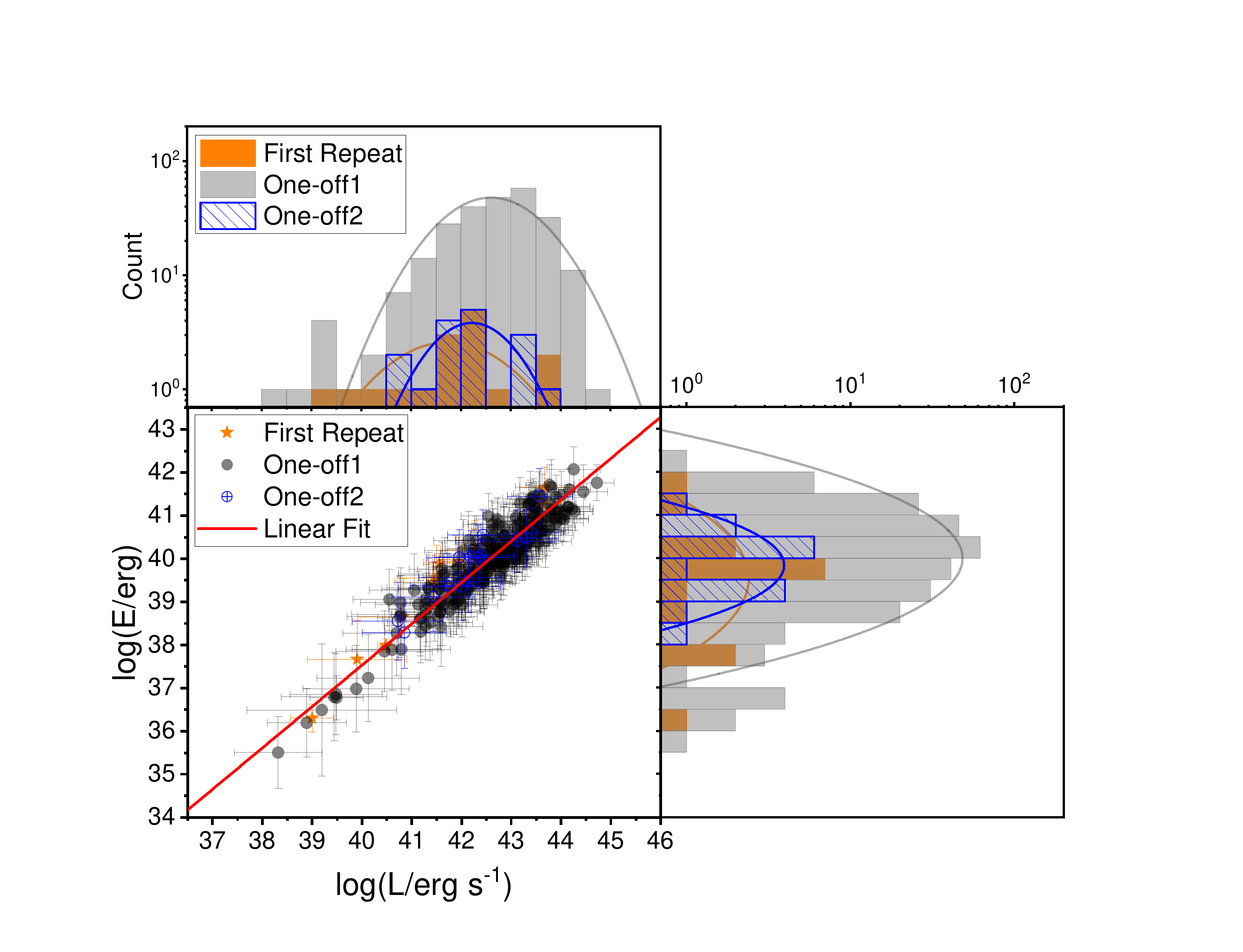}
	\caption{Scatter map comparing the peak luminosity and energy, as well as their distributions for those bursts with a well-determined peak frequency.}
	\label{fig:logLE}
\end{figure}

To obtain the peak luminosity ($L_{\rm p}$) and energy ($E$) of an FRB source, one needs to know its distance. For an FRB source with host galaxy identification like the case FRB 20121102A \citep{ten17}, 20180916B \citep{marc20}, or 20181030A \citep{bhar21b}, one can directly use its measured redshift to calculate its distance. For a source without host identification, one can estimate its redshift and distance using its extragalactic dispersion measure DM$_{\rm E}$ given by
\begin{equation}
	{\rm DM_{E}}={\rm DM_{cos}}+\frac{{\rm DM_{host}}}{1+z},
	\label{eq:DM_E}
\end{equation}
after excluding the contributions of the Milk Way interstellar medium (ISM) DM$_{\rm NE2001}$ 
(or DM$_{\rm YMW16}$)\footnote{We use DM$_{\rm YMW16}$.} 
and the Milky Way halo DM$_{\rm halo}$,
where ${\rm DM_{halo}}=50~{\rm pc~cm^{-3}}$ \citep{proch19} 
and the host galaxy contribution ${\rm DM_{host}}=129^{+66}_{-48}~{\rm pc~cm^{-3}}$ 
\citep{jam22a,jam22b} are adopted. The errors from the DM$_{\rm YMW16}$ and  ${\rm DM_{halo}}$ 
have been absorbed in that of the ${\rm DM_{host}}$ \citep{jam22a}.
The contribution from all extragalactic gas DM$_{\rm cos}$ in a flat $\Lambda$ cold dark matter universe is on average 
given by \citep{deng14,macq20}
\begin{equation}
	\left\langle\mathrm{DM}_{\text {cos }}\right\rangle=\int_{0}^{z_{\mathrm{FRB}}} \frac{c \bar{n}_{\mathrm{e}}(z) \mathrm{d} z}{H_{0}(1+z)^{2} \sqrt{\Omega_{\mathrm{m}}(1+z)^{3}+\Omega_{\Lambda}}},
	\label{eq:DM_cos}
\end{equation}
with mean density $\bar{n}_{\mathrm{e}}(z)=f_{\mathrm{d}}(z) \Omega_{\mathrm{b}} 3 H_{0}^{2} (8 \pi G)^{-1} (1+z)^{3} m_{\mathrm{p}}^{-1}\left(1-Y_{\mathrm{He}} / 2\right)$, where $H_0=67.4~{\rm km~s^{-1}~Mpc^{-1}}$ is the Hubble constant, 
and $\Omega_{\rm b}=0.044$, $\Omega_{\rm m}=0.315$, and $\Omega_{\Lambda}=0.685$ are the baryon, matter, 
and dark energy densities today \citep{planck20}, 
$m_{\rm p}$ is the proton mass and $Y_{\rm He}=0.25$ is the mass fraction of helium fully ionized in the gas, and
$f_{\mathrm{\rm d}}(z)$ is the fraction of cosmic baryons in diffuse ionized gas influenced by the redshift evolution of stars, stellar remnants, 
and the neutral ISM of galaxies\footnote{$f_{\rm d}(z)$ is calculated using the public FRB repository (https://github.com/FRBs/FRB) 
considering the redshift evolution of stars, stellar remnants, and the neutral ISM of galaxies \citep{macq20}.}. 
The uncertainty in DM$_{\rm cos}$ can be quantified as the fractional standard deviation $\sigma_{\rm DM}\approx Fz^{-0.5}$ for $z<1$ \citep{mcq14}, 
where $F$ quantifies the strength of the baryon feedback and is fixed as 0.32 based on \cite{macq20} and \cite{jam22a}.
From the Macquart relation \citep{macq20} and the results of \cite{jam22a}, an estimated FRB redshift $z$ from DM$_{\rm E}$ seems to be more uncertain 
if the DM$_{\rm E}$ is higher. Moreover, \cite{jam22a} obtained that the redshift $z$ even decreases against the increasing DM$_{\rm E}$ 
and argued that this reversal of the Macquart relation implies that 
it will vastly overestimate the FRB distance and thus energy from the very high DM$_{\rm E}$. 
This is supported by the case of FRB 190520B \citep{niu21}.
Therefore, we cut the FRB sources with ${\rm DM_{E}}\gtrsim 1500~{\rm pc~cm^{-3}}$ 
as we use the monotonically ascending ${\rm DM}-z$ relation from Equations (\ref{eq:DM_E}) and (\ref{eq:DM_cos}) 
to estimate the redshift and corresponding distance and their errors.
By using the redshift, distance, observed peak flux, fluence, and peak frequency of a burst, 
one can estimate its peak luminosity and energy from \citep{zhang18}
\be
L_{\rm p}\simeq10^{42}~{\rm erg~s^{-1}}4\pi \left(\frac{D_{\rm L}}{10^{28}~{\rm cm}}\right)^2\left(\frac{F_{\nu,\rm p}}{{\rm Jy}}\right)\left(\frac{\nu_{\rm p}}{{\rm GHz}}\right),
\label{eq:L_p_o}
\ee
and
\be
E\simeq10^{39}~{\rm erg}\left(\frac{4\pi}{1+z}\right)\left(\frac{D_{\rm L}}{10^{28}~{\rm cm}}\right)^2\left(\frac{S_{\nu}}{{\rm Jy~ms}}\right)\left(\frac{\nu_{\rm p}}{{\rm GHz}}\right),
\label{eq:E_o}
\ee
respectively. Here we use the peak frequency treated as the characteristic emission frequency to replace the central frequency because of two reasons. First, a very large part of bursts have a highest frequency higher than the top of the CHIME band and/or a lowest frequency lower than the bottom, so their central frequencies cannot be well estimated. But the majority of bursts have a well-determined peak frequency, neither higher than the top of the CHIME band or lower than the bottom. Second, for those bursts with both a well-estimated central frequency and a well-determined peak frequency, their central frequencies are comparable to the corresponding peak frequencies [$\nu_{\rm c}/{\rm MHz}=(1.06\pm0.01)\times( \nu_{\rm p}/{\rm MHz})-(21.58\pm5.57)$, Pearson's correlation coefficient $r_{\rm P}=0.998$], 
as displayed in Figure \ref{fig:pcfreq}. By using the bursts with a well-determined peak frequency, 
one can obtain the distributions of the peak luminosity and energy, 
as well as their correlation, as exhibited in Figure \ref{fig:logLE}. One can see that the peak luminosity and energy are obviously positively correlated 
with ${\rm log}(E/{\rm erg})=(0.96\pm0.04)\times{\rm log}(L/{\rm erg~s^{-1}})-(0.83\pm1.90)$ and Pearson's correlation
coefficient $r_{\rm P}=0.941$. It is roughly consistent with $E\sim L_{\rm p}\tau'$ assuming that the peak luminosity can represent the time-averaged luminosity. This may imply that the intrinsic burst-frame duration $\tau'$ for each burst is comparable, though it is relatively scattered. 
Additionally, the peak luminosity tends to be larger for the one-off1 sample from the Gaussian function fitting results listed in Table \ref{tab:results}, 
compared with the first repeat sample. 
The $p$-values of the peak luminosity between the one-off1 and first repeat samples strongly support this difference. 
This result is in accord with that in \cite{luo20} for the FRBs collected from other radio telescopes.
But the distribution difference between the one-off events and repeaters is not clearly seen in the energy from the $p$-values.

\section{Can Discriminant Properties Arise from Beamed Emission?}
\label{sec:explanation}
If non-repeating and repeating FRBs originate from one population, the observational discriminant properties between them could result from selection effects
(or propagation effects\footnote{In this paper, we do not consider propagation effects.}), e.g., the beaming effect.
In the following, we will qualitatively study whether all distribution distinctions of the properties 
mentioned in Section \ref{sec:properties} can arise from the beamed emission.

In the beamed emission model, based on \cite{con20}, one-off events tend to have a shorter duration $\tau$ mapping to a smaller collimated beaming solid angle $\Omega$ and are produced by a change in particles with a higher bulk Lorentz factor $\gamma$ than repeaters, i.e.,
\begin{equation}
\tau=C\Omega=C\frac{1}{\gamma}\ ({\rm or}\ C\frac{1}{\gamma^2}),
\label{eq:tau}
\end{equation}
where $C=5~{\rm ms}~1~{\rm sr}^{-1}$ is the coefficient with an assumed value, $\Omega=1/\gamma$ or $1/\gamma^2$ corresponding to the particles being confined to a thin sheet
or a pencil beam \citep{con20}.
On this basis, we do a detailed comparison between the model inferences and the observations within two radiation mechanisms: 
the coherent curvature radiation and synchrotron maser emission.

\subsection{Coherent Curvature Radiation}
\label{subsec:curvature}

\begin{figure}
	\centering
	\includegraphics[scale=0.55]{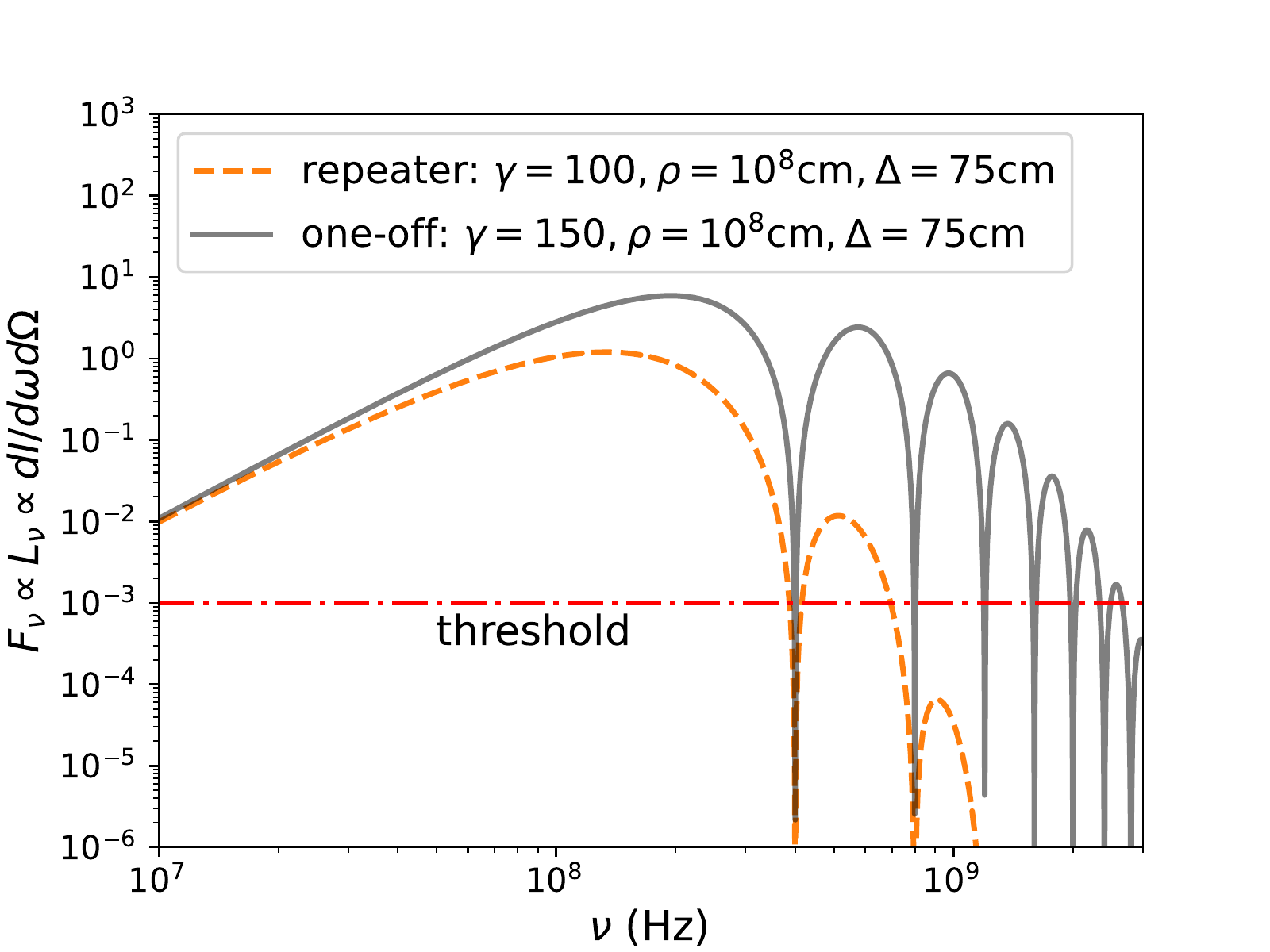}
	\caption{The coherent curvature radiation spectra in the bunches being two charge-separated clumps 
	with opposite signs for two different Lorentz factors (one-off: $\gamma=150$, and repeater: $\gamma=100$). 
	The red dashed-dotted line is an assumed telescope threshold.}
	\label{fig:Fv-v}
\end{figure}

{\em Bandwidth, spectral index, and peak frequency.} Within the coherent curvature radiation, an extremely steep spectrum can emerge only in the spectral oscillation \citep{yang18} or in the bunches consisting of two charge-separated clumps with opposite signs \citep{yang20}. Considering this type of bunches also with an oscillation, one can acquire the spectrum via \citep{yang20}
\be
\frac{dI_{(N)}}{d\omega d\Omega}=2\left[1-\cos\left(\frac{\omega\bm{n}\cdot\Delta}{c}\right)\right]N^2\frac{dI_{(1)}}{d\omega d\Omega}, 
\label{eq:dI_N}
\ee
where 
\be
\frac{dI_{(1)}}{d\omega d\Omega}\simeq\frac{e^2}{c}\left[\frac{\Gamma(2/3)}{\pi}\right]^2\left(\frac{3}{4}\right)^{1/3}\left(\frac{\omega\rho}{c}\right)^{2/3}e^{-\omega/\omega_{\rm c}},
\label{eq:dI_1}
\ee
and $\Delta$ is the separation between two clumps and the critical angular frequency $\omega_{\rm c}=2\pi\nu_{\rm c}=3\gamma^3c/2\rho$. 
If the difference in the bandwidth distributions between the one-off events and repeaters is just due to the beamed emission, 
i.e., Lorentz factor $\gamma$, by giving a set of parameters $\rho=10^8~{\rm cm}$, $\Delta=$ 75 cm, and $\gamma=150$ 
for a one-off event and $\gamma=100$ for a repeater, 
using \citep[see Equation (48) of][]{yang18}
\be
F_{\nu}=\frac{L_{\nu}}{4\pi D_{\rm L}^2}\propto\frac{dI_{(N)}}{d\omega d\Omega},
\label{eq:Fv}
\ee
and combining Equations (\ref{eq:dI_N}) and (\ref{eq:dI_1}), one can obtain the observed flux density spectra of the one-off event and repeater 
whose second components exactly occupy the CHIME band.
We can see that the spectrum of the one-off event has the same shape as the repeater but is steeper (larger positive and smaller negative spectral index) than the repeater either in the initial component or the subsequent oscillation, as displayed in Figure \ref{fig:Fv-v}. But interestingly, the steeper spectrum of the one-off event could result in a broader bandwidth than the repeater, given an assumed telescope threshold. Moreover, the peak frequency for the one-off event is seemingly higher than the repeater. These results seem in striking contrast to the statistical distribution 
differences between the one-off1 and first repeat samples 
for the spectral index and peak frequency, 
but in accordance with the distribution difference for the bandwidth
discussed in Sections \ref{subsec:duration} and \ref{subsec:sp_idx}.

\begin{figure}
	\centering
	\includegraphics[scale=0.55]{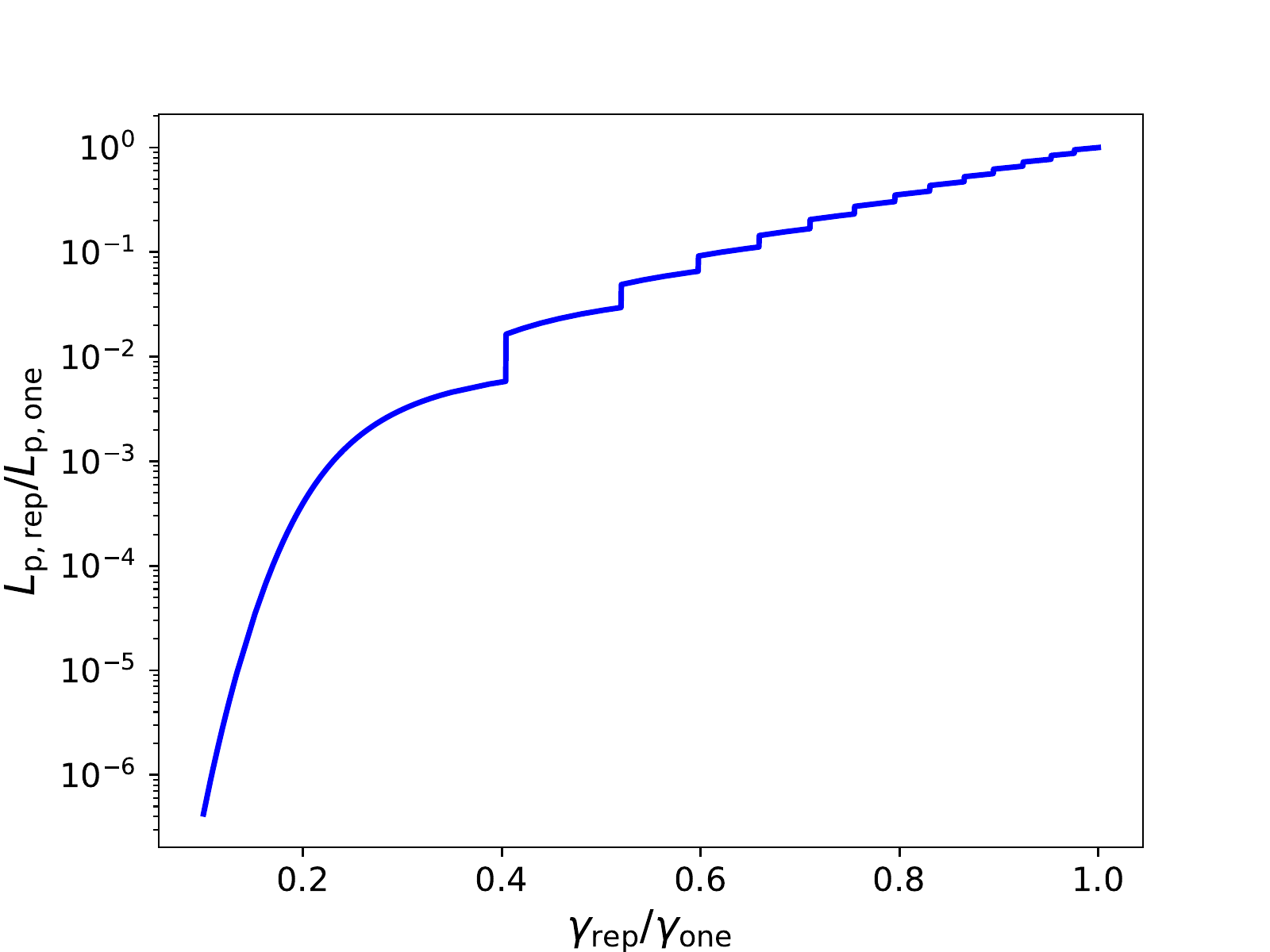}
	\includegraphics[scale=0.55]{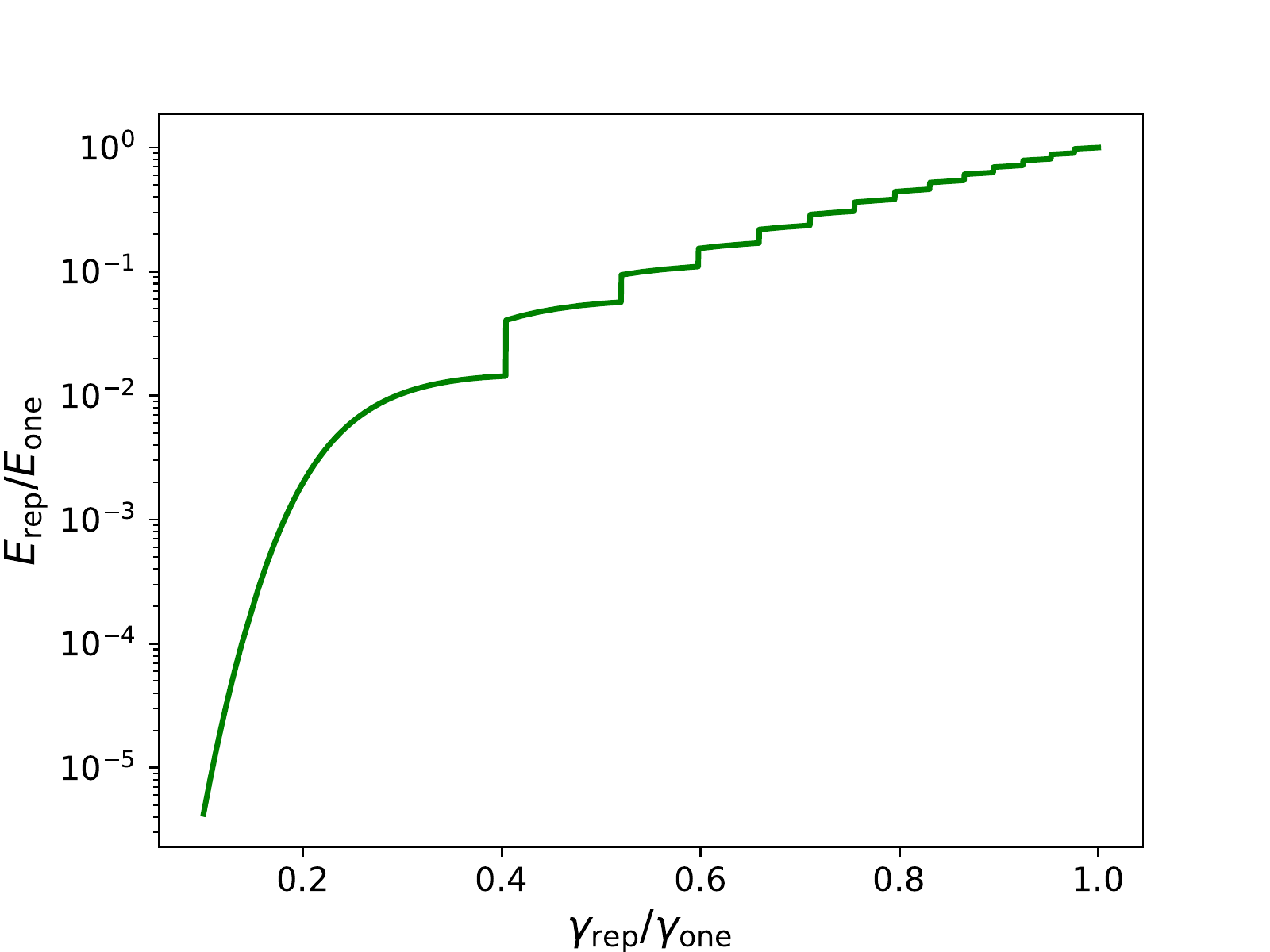}
	\caption{Upper panel: the peak luminosity ratios $L_{\rm rep}/L_{\rm one}$ vs. the Lorentz factor ratios $\gamma_{\rm rep}/\gamma_{\rm one}$ between a repeater and a one-off event. Lower panel: similar to the upper panel but for the energy.}
	\label{fig:LEratio}
\end{figure}

{\em Peak luminosity and energy.} Referring to Equations (\ref{eq:L_p_o}) and (\ref{eq:Fv}), one yields the peak luminosity
\be
L_{\rm p}\sim L_{\nu,\rm p}\nu_{\rm p}\propto\left.\frac{dI_{(N)}}{d\omega d\Omega}\right|_{\max}\times\nu_{\rm p}.
\label{eq:L_p_c}
\ee
Combining Equations (\ref{eq:dI_N}), (\ref{eq:dI_1}), and (\ref{eq:L_p_c}), 
one can numerically calculate the peak frequency $\nu_{\rm p}$ of the spectra and their corresponding peak flux density,
and thus the peak luminosity ratios for different Lorentz factor ratios between a one-off event and a repeater $\gamma_{\rm rep}/\gamma_{\rm one}$. 
The result is given in the upper panel of Figure \ref{fig:LEratio}. It is clear that the one-off event with 
a larger Lorentz factor compared with the repeater has a higher peak luminosity. 
This is in line with the statistical distribution distinction in luminosity 
between the one-off1 and first repeat samples discussed in Section \ref{subsec:luminosity}.

For the energy, it is roughly expressed as
\be
E\sim L_{\rm p}\tau,
\label{eq:E_c}
\ee
according to Equations (\ref{eq:L_p_o}) and (\ref{eq:E_o}).
Utilizing Equations (\ref{eq:tau}) and (\ref{eq:L_p_c}), one can get the energy ratios for different Lorentz factor ratios 
between a one-off event and a repeater $\gamma_{\rm rep}/\gamma_{\rm one}$ as displayed in the lower panel of Figure \ref{fig:LEratio}. 
This result seems inconsistent with the statistical distribution distinction in energy.

\subsection{Synchrotron Maser Emission}
\label{subsec:maser}
{\em Bandwidth, spectral index, and peak frequency.} In the framework of the synchrotron maser emission from decelerating relativistic blast waves \citep{met19}, the bulk Lorentz factor $\gamma$ in Equation (\ref{eq:tau}) stands for the Lorentz factor of the shocked gas toward the observer. For a synchrotron maser emission in the shocked gas frame,  
its spectral width, i.e., bandwidth, relates to the peak frequency of spectrum with $\Delta\nu_{\rm b}^{'}\lesssim(1-3)\nu_{\rm p}^{'}$ at magnetization $\sigma\gtrsim1$ \citep{plot19}.
In this case, the peak frequency and bandwidth in the observer frame are related to those in the shocked gas frame through
\begin{equation}
	\nu_{\rm p}=\gamma\nu_{\rm p}^{\prime}=3\gamma\nu_{\rm pm}\propto \tau^{-1/2}\propto \Omega^{-1/2},
	\label{eq:nu_p}
\end{equation}
and
\begin{equation}
	\Delta \nu_{\rm b}=\gamma\Delta\nu_{\rm b}^{\prime}\lesssim(1-3)\gamma\nu_{\rm p}^{\prime}\propto \tau^{-1/2}\propto \Omega^{-1/2},
	\label{eq:Delta_nu_b}
\end{equation}
where $\nu_{\rm pm}$ is the plasma frequency of the medium ahead of the shock.
This can explain the one-off events statistically have a broader bandwidth than the repeaters 
just due to the relativistic Doppler shift if they have the same intrinsic $\nu_{\rm p}^{\prime}$ and $\Delta\nu_{\rm b}^{\prime}$. 
In addition, $\Delta \nu_{\rm b}\lesssim(1-3)\nu_{\rm p}$ is consistent with 
the observed $\Delta \nu_{\rm b}/\nu_{\rm p}\lesssim$ 0.6 for the repeaters 
and $\Delta \nu_{\rm b}/\nu_{\rm p}\lesssim 1$ for the one-off events, as shown in the lower panels of Figure \ref{fig:sp_idx}. 
More importantly,
the spectrum from the synchrotron maser has a very sharp rising phase before the peak frequency because of the low-frequency cutoff 
and a relatively slow falling phase after the peak frequency \citep{plot19}, 
which seems to be in good agreement with the spectral index distributions lacking a very large negative value for 
either the repeaters or the one-off events. Furthermore, the distribution difference of the spectral index between the one-off events 
and repeaters in Section \ref{subsec:sp_idx} may indicate that the peak frequency and the rising phase of 
the spectrum for the repeaters are statistically located at the CHIME band, while for the one-off events, 
the peak frequency is usually below or close to the bottom of the CHIME band and the falling phase of the spectrum drops into the CHIME band. 
However, Equation (\ref{eq:nu_p}) implies that the one-off events with a larger Lorentz factor 
should have a higher observed peak frequency than the repeaters, 
being opposite to the statistical distribution results for the peak frequency mentioned in Section \ref{subsec:sp_idx}. 
This model inference for the peak frequency would further lead to the predicted distributions of 
the spectral index for the one-off events and repeaters seemingly being opposite to the statistical distribution results.

{\em Peak luminosity and energy.}
From Equation (12) of \cite{met19}, the peak luminosity and energy are given by
\begin{equation}
L_{\rm p}\approx4\pi R^2 n_{\rm ext}\gamma^4 m_{\rm p}c^3\propto\gamma^4,
\label{eq:L_p_s}
\end{equation}
and
\begin{equation}
E\sim L_{\rm p}\tau\approx 2\pi R^3n_{\rm ext}\gamma^2m_{\rm p}c^2\propto\gamma^2,
\label{eq:E_s}
\end{equation}
respectively, where $\tau=\frac{R}{2\gamma^{2}c}$ is used. These can give rise to the one-off events 
averagely having a higher peak luminosity 
and energy than the repeaters, consistent with the statistical distributions between the one-off1
and first repeater samples for the luminosity but not for the energy, 
analogous to the scenario in the coherent curvature radiation. 

\section{One Population or Two Populations?}
\label{sec:discussion}
From the morphology analysis in the first CHIME/FRB catalog, there are observed distribution distinctions in duration, bandwidth, 
spectral index, peak luminosity, and potentially peak frequency 
between the single-pulse once-off events and repeaters mentioned in Section \ref{sec:properties}. 
As explored in Section \ref{sec:explanation}, however, one population due to the beaming effect cannot well account for 
all distribution distinctions of properties between the single-pulse one-off events and repeaters. 
What is more, the beaming effect even can lead to the distributions of the peak frequency and spectral index for the one-off events 
and repeaters seemingly being opposite to the statistical distributions from the observations.
Therefore, this implies that there could be two populations for FRBs.
But what are the two populations?

\subsection{Are All FRBs Generated from Magnetars with Different Active Levels?}
The first possible scenario is that all FRBs are generated from magnetars, the population of non-repeating FRBs are from magnetars born in the delay formation channels \citep{mar19} such as neutron star$-$neutron star mergers associated with short gamma-ray bursts \citep{ross03,price06,giac13}, 
neutron star$-$white dwarf mergers possibly associated with rapidly evolving supernovae (SNe) \citep{too18,zhong20}, 
or white dwarf$-$white dwarf mergers \citep{yoon07,sch16} 
possibly associated with SNe Ia, and accretion-induced collapse of white dwarfs \citep{nomo91,tau13,sch15};
the population of repeating FRBs are from magnetars born in the prompt formation channels such as 
type I superluminous SNe, long gamma-ray bursts, or core-collapse SNe \citep{kasen10,duncan92,kou98}.
Nonetheless, it seems to be untrue based on the recent observations in host galaxies and offsets of the repeaters FRBs 20180916B \citep{marc20,ten21} 
and 20200120E \citep{bhar21a,kir21}. 
So it may be that the observed non-repeating and repeating FRBs 
are respectively from less active and more active magnetars due to different magnetic field strengths.
If this is the case, however, it is unclear whether the distribution discrepancies of properties between the one-off events and repeaters can arise 
from different magnetic field strengths. 
Moreover, if the one-off events and repeaters are produced by an identical mechanism like 
that generating FRB 20200428 and its associated X-ray burst, 
it is conceivable that the more active magnetars yielding repeaters could produce more frequent more violent outbreaks resulting in brighter bursts
than the less active magnetars yielding one-off events. This appears to be inconsistent with the statistical results shown in Figure \ref{fig:logLE}, 
though it is still a complicated issue.

 \begin{figure}
	\centering
	\includegraphics[scale=0.33]{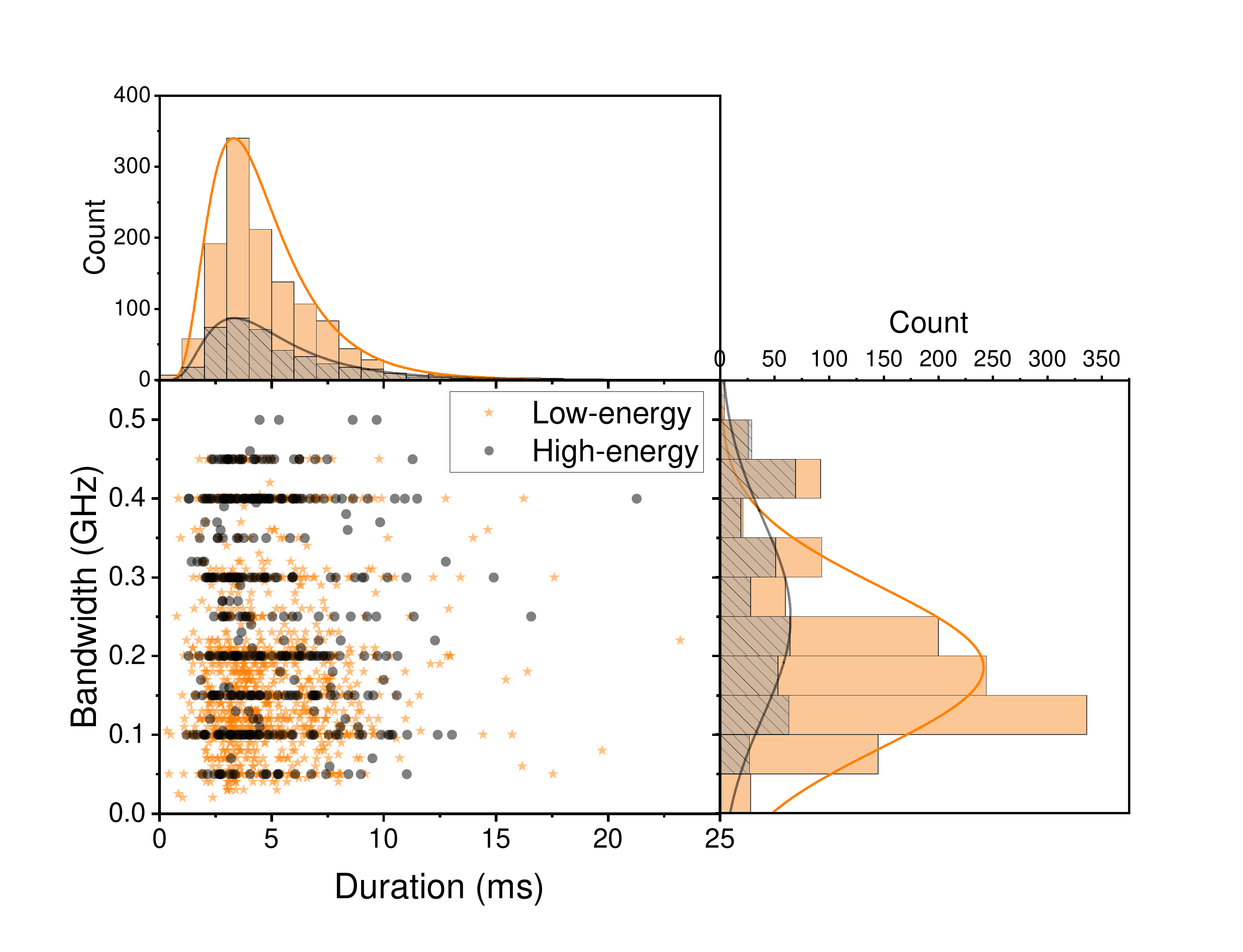}
	\caption{Distributions in duration (left panel) and bandwidth (right panel) for the low-energy bursts and high-energy bursts of FRB 20121102A 
		reported in \cite{lidi21}.}
	\label{fig:121102}
\end{figure}

\subsection{Are All FRBs Generated from Magnetars with Different Generation Mechanisms?}
The second possible scenario is that all FRBs are generated from magnetars, non-repeating and repeating 
FRBs are generated from different mechanisms.
Bring to mind the frequent short X-ray bursts and the infrequent giant flares from Galactic magnetars, 
they have distribution distinctions 
in peak frequency, peak luminosity, energy, etc \citep{tur15,kaspi17}. 
Note that these distinctions between the short bursts and the giant flares are across several orders of magnitude, 
whereas the distribution distinctions of properties between the one-off events and repeaters are just over 
a few times or one order of magnitude.
Therefore, those mechanisms giving rise to short X-ray bursts and giant flares from magnetars should not generate 
the distribution discrepancies between the one-off events and repeaters.
Also bring to mind the bimodal burst energy distribution
of FRB 20121102A recently reported by \cite{lidi21}\footnote{It should be noted that the bimodality in 
the energy distribution disappears when burst bandwidth, instead of the center frequency, is used to estimate energy \citep{agg21}.}, 
the infrequent high-energy bursts and frequent low-energy ones divided by energy ${\rm log}(E/{\rm erg})\sim38.2$ 
may be from two currently unknown different generation mechanisms. 
The distribution distinctions of properties between the one-off events and repeaters may also result from the two mechanisms. 
However, the high-energy and low-energy bursts of FRB 20121102A, 
as exhibited in Figure \ref{fig:121102}, do not show the same striking distribution distinction in duration ($p_{\rm KS}=0.15$, $p_{\rm AD}=0.05$)
though not in bandwidth ($p_{\rm KS}=3e-18$, $p_{\rm AD}<1e-3$) as the one-off events and repeaters in \cite{ple21}
\footnote{Although the Arecibo data of FRB 20121102A seem to show that low-energy bursts possess a more
typical long duration and narrow bandwidth than the high-energy bursts \citep{hew21}.}. 
This indicates that the two mechanisms producing the one-off events and repeaters could be different from 
those mechanisms producing the bimodal burst energy distribution of FRB 20121102A.
Hence, the two different mechanisms are still unknown and whether they can both occur in one magnetar is also still unknown .

\subsection{Are Repeating FRBs Generated from Magnetars, While Non-repeating FRBs Are from Catastrophic Events?}
A third possible scenario is that repeating FRBs are generated from magnetars,
while non-repeating FRBs are from catastrophic events likely involved in compact object mergers \citep{tota13,wang16,srid21} 
or the collapse of compact objects \citep{fal14,zhang14}
or others \citep{pla19},
though it is very likely that some one-off events observed by the CHIME telescope are 
actually repeater {\em wanglers} such as FRB 20121102A \citep{jose19}.
It is naturally unnecessary to consider the repetition rate question between repeating and non-repeating FRBs in this scenario.
There are many models that can give rise to non-repeating FRBs. 
Anyhow, they should possess a few discriminant features in the bursting process such as 
shorter timescale, etc. \citep{srid21}, in comparison to the magnetar model explaining repeating FRBs.

\section{Conclusions}
\label{sec:conclusions}
We have firstly made a comprehensive statistical analysis using the first CHIME/FRB catalog 
and identified a few distinct properties between the one-off events and repeaters,
and subsequently probed the question as to whether FRBs are classified into one or two populations, and finally obtained several meaningful results as follows.
\begin{itemize}
\item Between the single-pulse one-off events and repeaters in the catalog, 
there are a few distribution distinctions in duration, bandwidth, spectral index, peak luminosity, 
and potentially peak frequency for the complete samples. For example, the single-pulse
one-off events on average possess a shorter duration, broader bandwidth, smaller spectral index, lower peak frequency, 
and higher peak luminosity than the repeaters. 
In addition, the multi-pulse one-off events have the same distributions in bandwidth and spectral index
as the single-pulse one-off events for the complete samples, 
offset from the repeaters. But for the peak frequency, peak luminosity, or energy, 
there is no strong evidence for a distribution difference between the multi-pulse one-off events and repeaters.
\item Assuming that the one-off events and repeaters belong to one population, 
the statistical distribution distinctions
not in all properties can be well interpreted by the selection effect due to a beaming angle within 
either the coherent curvature radiation or the synchrotron maser emission mechanism by qualitative model analysis.
In particular, the distribution differences between the single-pulse one-off events and repeaters in the spectral index and peak frequency 
are seemingly opposite to the predictions of the beamed emission model. This implies that there could be two populations of FRBs.
\item What are the two populations? The first possible scenario is that all FRBs are generated from magnetars but with different activity levels. 
Non-repeating and repeating FRBs are respectively generated from less active and highly active magnetars due to different magnetic field strengths. 
However, it is unclear that whether the distribution discrepancies of properties between the one-off events and repeaters can arise 
from different magnetic field strengths. 
The second possible scenario is that all FRBs are generated from magnetars, and non-repeating and repeating FRBs are from different generation mechanisms.
But the two different generation mechanisms are unknown and whether they can both occur in one magnetar is also unknown.
In reality, it is more likely that the combination of the first and second scenarios results in the observational discriminant properties 
between the one-off events and repeaters.
The third possible scenario is repeating FRBs are generated from magnetars,
while non-repeating FRBs are from catastrophic events likely involved in compact object mergers or the collapse of compact objects or others. But these models of catastrophic events should possess a few discriminant features in the bursting process compared with the magnetar model accounting for repeating FRBs to satisfy the observed discriminant properties between the one-off events and repeaters.
\end{itemize}

\acknowledgments
We thank the referee for helpful
comments and suggestions that have allowed us to significantly improve the
presentation of this manuscript. We also thank Ziggy Pleunis for helpful discussions on the 
first CHIME/FRB catalog.
S.Q.Z. acknowledges support from the China Postdoctoral Science Foundation (grant No. 2021TQ0325). 
W.J.X. is supported by the Program of China Scholarships Council (grant No. 202006660005).
C.M.D. is supported by the Guangxi Science Foundation (grant No. 2021AC19263).
Z.G.D. is supported by the National Key
Research and Development Program of China (grant No.
2017YFA0402600), National SKA Program of China (grant No. 2020SKA0120300), 
and National Natural Science Foundation
of China (grant No. 11833003). 


\end{document}